\documentclass[conference]{IEEEtran}

\IEEEoverridecommandlockouts

\usepackage{cite}
\usepackage{amsmath}

\usepackage{marvosym}

\usepackage{caption}

\usepackage{bbm}
\usepackage{bm}
\usepackage{amssymb}
\usepackage{graphicx, subfig}
\usepackage{wrapfig}

\usepackage[ruled,vlined]{algorithm2e}

\usepackage{amsmath,amssymb,amsfonts}
\usepackage{textcomp}
\usepackage{xcolor}
\usepackage{setspace}

\usepackage[marginal]{footmisc}

\usepackage[noend]{algpseudocode}

\usepackage{url}

\def\BibTeX{{\rm B\kern-.05em{\sc i\kern-.025em b}\kern-.08em
    T\kern-.1667em\lower.7ex\hbox{E}\kern-.125emX}}

\begin{document}

\title{Team Channel-SLAM: A Cooperative Mapping Approach to Vehicle Localization\\
}

\author{
\IEEEauthorblockN{Xinghe Chu$^{1,2,4}$,
   \Letter Zhaoming Lu$^{1,2,4}$, Luhan Wang$^{1,2,4,5}$, Xiangming Wen$^{1,2,4}$, David Gesbert$^{3}$}

\IEEEauthorblockA{${}^1$Beijing University of Posts and Telecommunications, Beijing, China}

\IEEEauthorblockA{${}^2$Beijing Laboratory of Advanced Information Networks, Beijing, China}

\IEEEauthorblockA{${}^3$Communications Systems Department, EURECOM, Sophia Antipolis, France}

\IEEEauthorblockA{${}^4$Beijing Key Laboratory of Network System Architecture and Convergence, Beijing, China}

\IEEEauthorblockA{${}^5$Witcomm Open Source Communication Research Institute, Beijing, China}

}

\maketitle

\begin{abstract}

Vehicle positioning is considered a key element in autonomous driving systems. While conventional positioning requires the use of GPS and/or beacon signals from network infrastructure for triangulation, they are sensitive to multi-path and signal obstruction. However, recent proposals like the Channel-SLAM method showed it was possible in principle to in fact leverage multi-path to improve positioning of a single vehicle.
In this paper, we derive a cooperative Channel-SLAM framework, which is referred as Team Channel-SLAM. Different from the previous work, Team Channel-SLAM not only exploits the stationary nature of reflecting objects around the receiver to characterize the location of a single vehicle through multi-path signals, but also capitalizes on the multi-vehicle aspects of road traffic to further improve positioning.
Specifically, Team Channel-SLAM exploits the correlation between reflectors around multiple neighboring vehicles to achieve high precision multiple vehicle positioning. Our method uses affinity propagation clustering and cooperative particle filter. The new framework is shown to give substantial improvement over the single vehicle positioning situation.

\end{abstract}


\section{Introduction}

Positioning is widely considered to play a key role in vehicular navigation and autonomous driving \cite{kuutti2018survey}. However, the typical positioning approaches like GPS, IMU and LiDAR can be limited by obstruction, error accumulation or severe weather conditions \cite{kuutti2018survey}.
The development of 5G has improved the time resolution and angle resolution greatly \cite{3gpp.38.211}, which brings new opportunities for radio based localization and navigation.

There are already many interesting contributions to the radio based positioning problems \cite{del2017survey,win2018theoretical,Marano2010NLOS ,koivisto2017joint}. However, multi-path propagation has long been regarded as a drawback for radio based localization technologies. Hence many contributions focus on multi-path elimination and LOS path extraction \cite{horiba2015accurate,jiao2009lcrt}. But recent research found that multi-paths can bring additional information for localization, which makes it possible to turn them from foe to friend in the context of radio based localization \cite{witrisal2016high}. In this context, \cite{wang2011omnidirectional} presents a NLOS identification and localization scheme and \cite{han2018hidden} presents an algorithm to estimate the position and the size of vehicle by using multi-paths when LOS link is obstructed.
While the above methods are typically presented in the context of estimation from a single time-slot worth of data, they can easily be extended to multiple time slot processing as some features (like reflecting planes) remain static over multiple time slots, which will further enhance positioning in the presence of noise.

As another powerful alternative, the so-called Channel-SLAM based method was proposed \cite{gentner2016multipath}. This technique estimates the user's position with a simultaneous localization and mapping method. There has been several Channel-SLAM based variants in the literature, but all of them are aiming at single vehicle localization \cite{leitinger2019belief,yassin2018mosaic,mendrzik2018joint,palacios2017jade}.
  However, as a crucial point raised in this paper,
in moderately dense traffic or even normal traffic scenario, we can expect groups of 2 or 3 vehicles to closely follow each other, leading to a situation where these vehicles will share at least one or two key reflectors. These reflectors can be multiple observed by vehicles in different positions, so that they can be estimated through multiple observations from both time domain and space domain.

In this paper, a new method referred to as Team Channel-SLAM is proposed to exploit the multi-vehicle nature of typical road traffic and the ability for different vehicles to cooperate with each other so as to improve the localization performance. The key point behind this contribution is the recognition that the existence of several common reflectors provide additional {\em structure} to the multiple vehicle localization problem, hence leading to improved estimation performance. In the Channel-SLAM algorithm, the reflectors are associated with the existence of virtual transmitters (VT) from which reflections {\it appear} to originate from. In our new framework, the VT corresponding to reflectors that are shared by multiple vehicles are merged together, giving rise to {\it common virtual transmitters} (CVT). Affinity propagation based clustering method is used for CVT cluster formation. Then CVT particle filters and vehicle particle filters are used to locate and track the position of vehicles cooperatively, where the former is used for CVT positioning and the latter is used for vehicle localization. Numerical results show that Team Channel-SLAM leads to over $40\%$ improvement over single vehicle situation when traffic density as low as 4 vehicles per 140 meters in an 8-lane road.

\section{System Model}

\subsection{State Model}
We consider a scenario with one base station (BS) and $M$ vehicles. Each vehicle is equipped with one radio user equipment.
The state of $m$-th vehicle at time ${t_k}$ is denoted as
\begin{equation}
 {X_{{V_m}}}\left( {{t_k}} \right) = \left\{ {{r_{{V_m}}}\left( {{t_k}} \right),{v_{{V_m}}}\left( {{t_k}} \right)} \right\}\label{state_m_veh}
\end{equation}
where ${r_{{V_m}}}\left( {{t_k}} \right) \in {{\rm{R}}^2}$ and ${v_{{V_m}}}\left( {{t_k}} \right) \in {{\rm{R}}^2}$
are its position and velocity, respectively.

\begin{figure}[t]
\centerline{\includegraphics[width=0.4\textwidth]{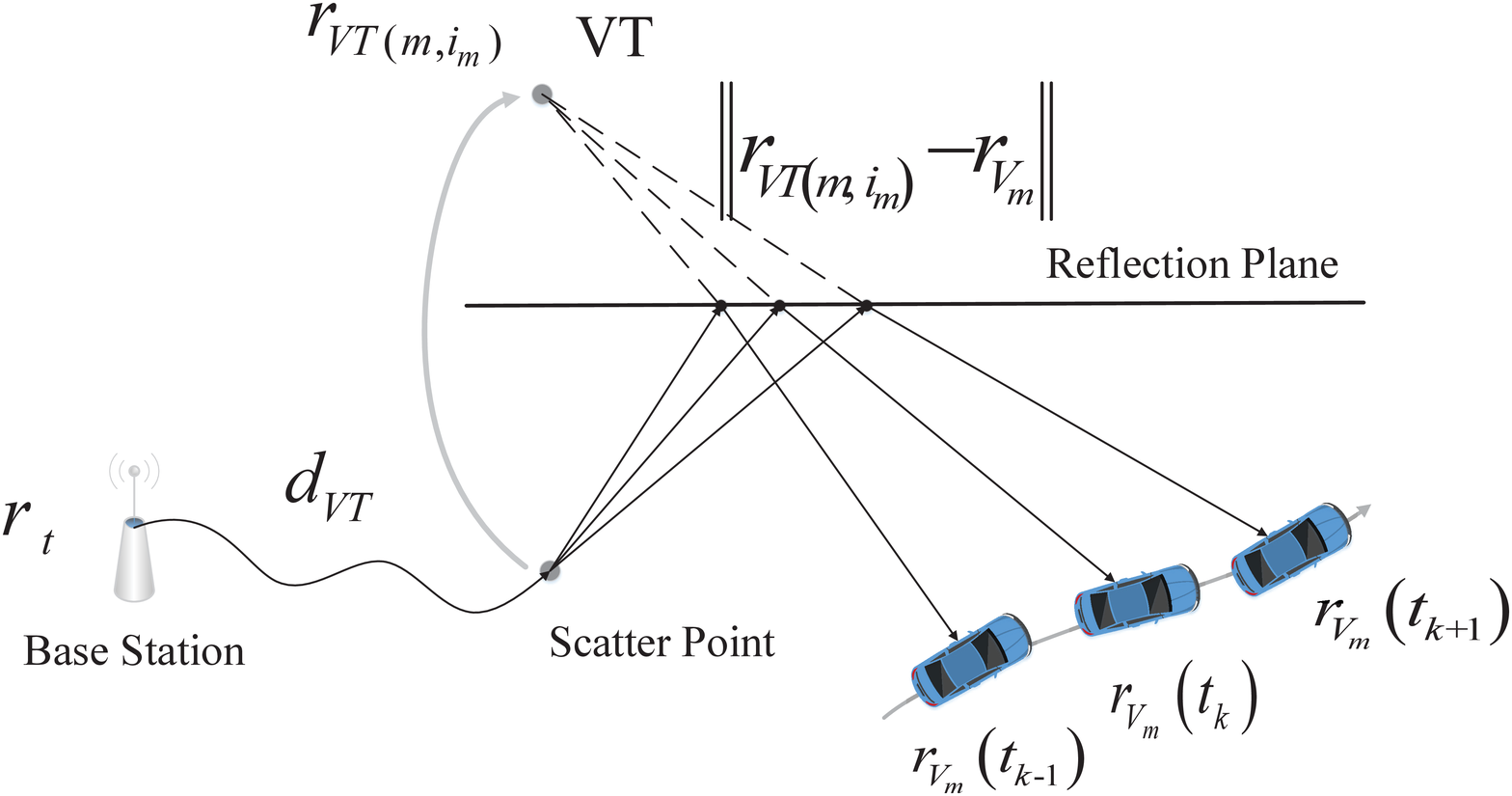}}
\caption{Signal from base station is scattered and then reflected to vehicle. In this situation, the mirror symmetry of the scatter point can be seen as a virtual transmitter originating LOS signal to the moving vehicle. ${d_{VT}}$ is the additional distance, which is larger than zero when scattering happens.}
\label{fig1}
\end{figure}

As shown in Fig. \ref{fig1}, the multi-path at time ${t_k}$ for vehicle $m$ can be modeled as a LOS link signal originating from a VT adding an additional distance, where the position of VT is constant over time \cite{gentner2016multipath}. So the state of VTs from vehicle $m$ can be denoted as
\begin{equation}
  {X_{V{T_m}}}({t_k}) = \left\{ {{r_{VT\left( {m,{i_m}} \right)}}\left( {{t_k}} \right),{d_{VT\left( {m,{i_m}} \right)}}\left( {{t_k}} \right)} \right\}_{{i_m} = 1}^{{N_m}\left( {{t_k}} \right)}\label{state_vt}
\end{equation}
where ${N_m}\left( {{t_k}} \right)$ is the number of multi-paths for vehicle $m$ at time slot ${t_k}$, and ${r_{VT\left( {m,{i_m}} \right)}}\left( {{t_k}} \right) \in {{\rm{R}}^{\rm{3}}},{d_{VT\left( {m,{i_m}} \right)}}\left( {{t_k}} \right) \in {{\rm{R}}}$ denote the position and additional distance for corresponding VT, respectively.

\subsection{Observation Model}
The observations from vehicles include Time of Arrival (ToA) and Angle of Arrival (AoA), which can be extracted from the signals received at the base station. The observation for vehicle $m$ at time $t_k$ is denoted as
\begin{equation}
  \begin{array}{l}
{Z_m}\left( {{t_k}} \right) = \left\{ {{{\hat \alpha }_{\left( {m,{i_m}} \right)}}\left( {{t_k}} \right),{{\hat d}_{\left( {m,{i_m}} \right)}}\left( {{t_k}} \right)} \right\}_{{i_m} = 1}^{{N_m}\left( {{t_k}} \right)}\\
{{\hat \alpha }_{\left( {m,{i_m}} \right)}}\left( {{t_k}} \right) = \left( {{{\hat \theta }_{\left( {m,{i_m}} \right)}},{{\hat \varphi }_{\left( {m,{i_m}} \right)}}} \right)
\end{array}\label{ob_state}
\end{equation}
where ${{{\widehat \theta }_{\left( {m,{i_m}} \right)}}}$ and ${{{\widehat \varphi }_{\left( {m,{i_m}} \right)}}}$ stand for the polar angle and azimuth angle of AoA estimation,
and ${{{\hat d}_{\left( {m,{i_m}} \right)}}\left( {{t_k}} \right)}$ is the ToA estimation multiplied by speed of light.
As mentioned in section II-A, the travel distance of a multi-path can be modeled as a LOS path distance from a VT plus an additional distance.
\begin{equation}
  \begin{array}{c}
  {r_{VT\left( {m,{i_m}} \right)}}\left( {{t_k}} \right) = {r_{{V_m}}}\left( {{t_k}} \right) + {\Delta _{\left( {m,{i_m}} \right)}}\left( {{t_k}} \right) \cdot \overrightarrow R \left( {{{\widehat \alpha }_{\left( {m,{i_m}} \right)}}\left( {{t_k}} \right)} \right) \vspace{0.5ex} \\
  {\Delta _{\left( {m,{i_m}} \right)}}\left( {{t_k}} \right) = {{\hat d}_{\left( {m,{i_m}} \right)}}\left( {{t_k}} \right) - {d_{VT\left( {m,{i_m}} \right)}}
  \end{array} \label{additional_dis_def}
\end{equation}
where $\vec R\left( {a,b} \right) = \left( {\cos a\sin b,\sin a\sin b,\cos b} \right)$ is the vector of polar angle $a$ and azimuth angle $b$. ${d_{VT\left( {m,{i_m}} \right)}}\left( {{t_k}} \right)$ is zero if there is no scattering, otherwise ${d_{VT\left( {m,{i_m}} \right)}}\left( {{t_k}} \right)$ is larger than zero.
\begin{figure}[tb]
\centerline{\includegraphics[width=0.295\textwidth]{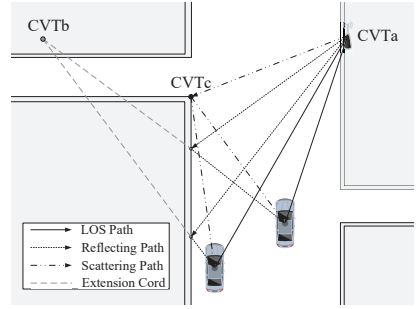}}
\caption{The model of three kinds of CVT is shown in this figure. CVT-a is the base station. CVT-b locates at the mirror position of base station. CVT-c locates at the scattering point.   }
\label{fig2}
\end{figure}
\section{CVT Cluster Formation}
As shown in Fig. \ref{fig2}, we denote the  number of VTs observed by vehicle $m$ at time $t_k$ as ${N_m}\left( {{t_k}} \right)$, so there is totally $\sum\limits_{m = 1}^M {{N_m}\left( {{t_k}} \right)}$ different VTs at time ${t_k}$. However, a group of VTs may find themselves in the same location
when those VTs are observed by: a) all the LOS links received by different vehicles, b) two or more multi-paths received by different vehicles with the same reflecting plane, c) two or more multi-paths received by different vehicles with the same scattering point. In this case, they can be treated as one common virtual transmitter (CVT).
 What's more, for situation b), if the Cartesian equations of two reflecting planes are the same, the VTs observed from multi-paths reflected by the two planes can also be seen as one CVT, for the position of them are also the same in theory.

However, those VTs treated as one CVT are usually not completely identical because the noise corrupts the observations. But they should be close with each other, so most likely there will be clusters of VTs in multiple vehicle scenario, where each cluster constitutes multiple observations for one CVT. This will be the role of the clustering algorithm to link together a group of neighboring VTs.

The key advantage of the above clustering technique is
that the position of CVTs can be then estimated on the basis a greater amount of the observations from the different neighboring vehicles.

Affinity propagation \cite{frey2007clustering,liu2017distributed} can cluster a group of nodes by choosing an exemplar (cluster head) for each node without preknowledge of cluster number and size.
The logarithm of the distance between VTs is defined as the {\it similarity value}.
\begin{equation}
  \begin{array}{*{20}{l}}
  {s\left( {p,q} \right) =  - \ln \left( {\left\| {{r_{V{T_p}}} - {r_{V{T_q}}}} \right\| + 1} \right),p \ne q} \vspace{0.5ex} \\
  {{r_{V{T_p}}},{r_{V{T_q}}} \in \left\{ {{r_{VT\left( {m,{i_m}} \right)}}} \right\}_{m\; = \;1,{i_m} = 1}^{m = M,{i_m} = {N_m}\left( {{t_k}} \right)}}
  \end{array}\label{ap}
\end{equation}
where $s\left( {p,q} \right)$ is the similarity value of VT $p$ with respect to VT $q$, which indicates how well VT $q$ is suited
to be the exemplar for VT $p$ \cite{frey2007clustering}.

The affinity propagation algorithm continuously updates and iterates a so-called {\it responsibility value} $r\left( {p,q} \right)$ and {\it availability value} $a\left( {p,q} \right)$, where the former means the accumulated evidence for how well-suited VT
$q$ can serve as the exemplar for VT $p$ and the latter means the accumulated evidence for how
appropriate it would be for VT $p$ to choose
VT $q$ as its exemplar \cite{frey2007clustering}.
 Finally, for VT $p$, the VT $q$ that maximizes the sum of responsibility value and availability value
is selected as an exemplar of VT $p$ \cite{frey2007clustering}. The iteration of affinity propagation for CVT cluster formation is done as follows:

\noindent \textbf{i)} Calculate the responsibility value:
\begin{equation}
r\left( {p,q} \right) \leftarrow s\left( {p,q} \right) - \mathop {\max }\limits_{q'{\rm{s}}.{\rm{t}}.q' \ne q} \left\{ {a\left( {p,q'} \right) + s\left( {p,q'} \right)} \right\} \label{eq1}
\end{equation}
\textbf{ii)} Damp the responsibility value:
\begin{equation}
r\left( {p,q} \right) \leftarrow \left( {1 - \lambda } \right)r\left( {p,q} \right) + r{\left( {p,q} \right)^{old}} \label{eq2}
\end{equation}
where $\lambda $ is damping factor between 0 and 1, and $r{\left( {p,q} \right)^{old}}$ is the responsibility value at previous iteration.

\noindent \textbf{iii)} Calculate the availability value:
\begin{equation}
a\left( {p,q} \right) \leftarrow \min \left\{ {0,r\left( {q,q} \right) + \sum\limits_{{p^\prime } \notin \left\{ {p,q} \right\}} {\max \left\{ {0,r\left( {p',q} \right)} \right\}} } \right\}\label{eq3}
\end{equation}
\textbf{iv)} Damp the availability value:
\begin{equation}
a\left( {p,q} \right) \leftarrow \left( {1 - \lambda } \right)a\left( {p,q} \right) + a{\left( {p,q} \right)^{old}} \label{eq4}
\end{equation}
\textbf{v)} Calculate the self-availability value:
\begin{equation}
a\left( {q,q} \right) \leftarrow \sum\limits_{{p^\prime } \ne q} {\max \left\{ {0,r\left( {p',q} \right)} \right\}} \label{eq5}
\end{equation}
\textbf{vi)} Choose exemplars:
\begin{equation}
  \begin{array}{*{20}{l}}
  {{\rm{if}}\;\;iter > {N_{iter}}\;\;{E_p} = \mathop {\arg \max }\limits_q \left\{ {a\left( {p,q} \right) + r\left( {p,q} \right)} \right\}}\\
  {{\rm{if}}\;\;iter \le {N_{iter}}\;\;{\rm{return}}\;\;\rm{i)}  }
  \end{array}\label{eq6}
\end{equation}
where ${E_p}$ is the exemplar for VT $p$.

After the above steps, exemplars are chosen for each VT and those VTs with the same exemplar will make up one CVT cluster.
Let ${N_{C}}\left( {{t_k}} \right)$ denote the number of clusters
 .
The state of CVT cluster is denoted as
\begin{equation}
{X_{C}}\left( {{t_k}} \right) = \left\{ {{X_{C_u}}\left( {{t_k}} \right)} \right\}_{u = 1}^{{N_{C}}\left( {{t_k}} \right)} \label{eq7}
\end{equation}
where ${X_{{C_u}}}\left( {{t_k}} \right) = \left\{ {\rm{CV}{{\rm{T}}_u}\left( {{t_k}} \right),{\rm{CVT}}{{\rm{I}}_u}\left( {{t_k}} \right)} \right\}$ is the state of CVT cluster $u$. ${\rm{CV}}{{\rm{T}}_u}\left( {{t_k}} \right) = \left\{ {{r_{{C_u}}}\left( {{t_k}} \right),{d_{{C_u}}}\left( {{t_k}} \right)} \right\}$ is the position and additional distance of CVT cluster $u$, respectively.
\begin{equation}
\begin{array}{*{20}{l}}
{{r_{{C_u}}}\left( {{t_k}} \right) = \mathbb{E}\left( {{r_{VT({i_u})}}\left( {{t_k}} \right)} \right),{i_u} = 1,2,...,{N_u}\left( {{t_k}} \right)}\\
{{d_{{C_u}}}\left( {{t_k}} \right) = \mathbb{E}\left( {{d_{VT({i_u})}}\left( {{t_k}} \right)} \right),{i_u} = 1,2,...,{N_u}\left( {{t_k}} \right)}
\end{array}\label{eq8}
\end{equation}
$\mathbb{E}\left(  \cdot  \right)$ is the expectation operator and ${{N_u}\left( {{t_k}} \right)}$ is the number of VT observations in CVT cluster $u$.
\begin{equation}
{\rm{CVT}}{{\rm{I}}_u}\left( {{t_k}} \right){\rm{ = }}\left[ {{\mathbbm{1}_{u,1}}\left( {{t_k}} \right),...,{\mathbbm{1}_{u,m}}\left( {{t_k}} \right),...,{\mathbbm{1}_{u,M}}\left( {{t_k}} \right)} \right]\vspace{0.5ex} \label{CVTI_def}
\end{equation}
${\rm{CVT}}{{\rm{I}}_u}\left( {{t_k}} \right)$ is the common virtual transmitter index that indicates whether vehicle $m$ observes a VT that belongs to CVT cluster $u$. If it does, its $m$-th element ${\mathbbm{1}_{u,m}}\left( {{t_k}} \right)$ equels to the corresponding multi-path index $i_m$ and otherwise ${\mathbbm{1}_{u,m}}\left( {{t_k}} \right)$ is zero.

\section{Cooperative Simultaneous Localization and Mapping}
A particle filter \cite{Siciliano2016Robotics} approach is described for multiple vehicle tracking and CVT positioning. CVT particle filters and vehicle particle filters are introduced to estimate the state of CVTs and vehicles jointly, which is shown in Algorithm 1.

\subsection{CVT Particle Filter}
The PDF of the $u$-th CVT can be calculated by
\begin{equation}
  \begin{array}{*{20}{l}}
  {{\rm{p}}\left( {{\rm{CV}}{{\rm{T}}_u}\left( {{t_k}} \right)\left| {{Z_{V\left( u \right)}}\left( {{t_k}} \right);{\Upsilon _{V(u)}} = {\Upsilon _{V(u)}}\left( {{t_k}} \right)} \right.} \right)}\\
  { \approx \sum\limits_{a = 1}^{{N_C}} {w_{{C_u}}^{(a)}\left( {{t_k}} \right) \times \delta \left( {{\rm{CV}}{{\rm{T}}_u}\left( {{t_k}} \right) - {\rm{CVT}}_u^{(a)}\left( {{t_k}} \right)} \right)} }
\end{array}\label{eq9}
\end{equation}
where ${V(u)}$ is the set of the vehicles that observe a VT belonging to CVT cluster $u$.
\begin{equation}
  V\left( u \right) = \left\{ {{m_u}\left| {{{\mathbbm{1}}_{u,{m_u}}} \ne 0} \right.} \right\}\label{v_u_define}
\end{equation}
${Z_{V\left( u \right)}}\left( {{t_k}} \right)$ denotes the observation from vehicles in ${V(u)}$. ${{\rm{CVT}}_u^{(a)}\left( {{t_k}} \right)}$ is the $a$-th particle of CVT state ${{\rm{CV}}{{\rm{T}}_u}\left( {{t_k}} \right)}$ and ${{\Upsilon _{V(u)}}\left( {{t_k}} \right)}$ is the vehicle particles in ${V(u)}$,

\begin{equation}
  {\Upsilon _{V(u)}}\left( {{t_k}} \right) = \left\{ {\left\{ {r_{V,{m_u}}^{(j)}({t_k})} \right\}_{j = 1}^{{N_V}}} \right\},{m_u} \in V\left( u \right)\label{parmic_veh2cvt}
\end{equation}
${w_{C_u}^{(a)}\left( {{t_k}} \right)}$ is the weight of each particle, the weight update euqation is denoted as:
\begin{equation}
  \begin{array}{*{20}{l}}
  {w_{{C_u}}^{(a)}\left( {{t_k}} \right) = w_{{C_u}}^{(a)}\left( {{t_{k - 1}}} \right)}\\
  { \times {\rm{p}}\left( {{Z_{V\left( u \right)}}\left( {{t_k}} \right)\left| {{\rm{CVT}}_u^{(a)}\left( {{t_k}} \right);{\Upsilon _{V(u)}} = {\Upsilon _{V(u)}}\left( {{t_k}} \right)} \right.} \right)}
\end{array}\label{cvt_weight_upd}
\end{equation}
The CVT particles can be drawn from the following distribution:
\begin{equation}
  \begin{array}{c}
  {\rm{CVT}}_u^{(a)}\left( {{t_k}} \right)\sim{\rm{p}}\left( {{\rm{CVT}}_u^{(a)}\left( {{t_k}} \right)\left| {{\rm{CVT}}_u^{(a)}\left( {{t_{k - 1}}} \right)} \right.} \right)\\
  {\rm{ = }}\delta \left( {{\rm{CVT}}_u^{(a)}\left( {{t_k}} \right) - {\rm{CVT}}_u^{(a)}\left( {{t_{k - 1}}} \right)} \right)
  \end{array}\label{eq_cvt_particle}
\end{equation}
\subsection{Vehicle Particle Filter}
The PDF of particles corresponding to the $m$-th vehicle can be calculated by
\begin{equation}
  \begin{array}{*{20}{l}}
  {{\rm{p}}\left( {{r_{{V_m}}}\left( {{t_k}} \right)\left| {{Z_m}\left( {{t_k}} \right),{U_m}\left( {{t_k}} \right);{\Upsilon _{C\left( m \right)}} = {\Upsilon _{C\left( m \right)}}\left( {{t_k}} \right)} \right.} \right)}\\
  { \approx \sum\limits_{j = 1}^{{N_V}} {w_{{V_m}}^{(j)}\left( {{t_k}} \right)\delta \left( {{r_{{V_m}}}\left( {{t_k}} \right) - r_{{V_m}}^{\left( j \right)}\left( {{t_k}} \right)} \right)} }
\end{array} \label{eq10}
\end{equation}
where ${{Z_m}\left( {{t_k}} \right)}$ is the observation from vehicle $m$ at time ${t_k}$, ${{U_m}\left( {{t_k}} \right)}$ is the motion information for vehicle $m$ at time ${t_k}$, ${C(m)}$ is the set of the CVT clusters that contain a VT observed by vehicle $m$.
\begin{equation}
  C\left( m \right) = \left\{ {{u_m}\left| {{{\mathbbm{1}}_{{u_m},m}} \ne 0} \right.} \right\}\label{c_m_define}
\end{equation}
${\Upsilon _{C\left( m \right)}}\left( {{t_k}} \right)$ is the CVT particles in ${C(m)}$,
\begin{equation}
{\Upsilon _{C\left( m \right)}}\left( {{t_k}} \right) = \left\{ {\left\{ {{\rm{CVT}}_{{u_m}}^{(a)}({t_k})} \right\}_{a = 1}^{{N_C}}} \right\},{u_m} \in C\left( m \right)\label{parmic_cvt2veh}
\end{equation}
${w_{{V_m}}^{(j)}\left( {{t_k}} \right)}$ is the weight of each vehicle particle, the weight update equation is denoted as:
\begin{equation}
  \begin{array}{*{20}{l}}
  {w_{{V_m}}^{(j)}\left( {{t_k}} \right) = w_{{V_m}}^{(j)}\left( {{t_{k - 1}}} \right)}\\
  { \times {\rm{p}}\left( {{Z_m}\left( {{t_k}} \right)\left| {r_{{V_m}}^{(j)}\left( {{t_k}} \right)} \right.;{\Upsilon _{C\left( m \right)}} = {\Upsilon _{C\left( m \right)}}\left( {{t_k}} \right)} \right)}
\end{array}\label{eq_veh_weight_upd}
\end{equation}
The vehicle particles can be drawn from the following distribution:
\begin{equation}
    r_{{V_m}}^{(j)}\left( {{t_k}} \right)\sim {\rm{p}}\left( {r_{{V_m}}^{(j)}\left( {{t_k}} \right)\left| {r_{{V_m}}^{(j)}\left( {{t_{k - 1}}} \right)} \right.,{U_m}\left( {{t_k}} \right)} \right)\label{eq_veh_particle}
\end{equation}

\subsection{Vehicle Motion Update}
In the scenario, suppose the error of motion information follows Gaussian distribution. The motion information for vehicle $m$ at time $t_k$ can be denoted as
\begin{equation}
  \begin{array}{*{20}{c}}
  {{U_m}\left( {{t_k}} \right) = \left\{ {{v_{{V_m}}}\left( {{t_{k{\rm{ - 1}}}}} \right),{v_{{V_m}}}\left( {{t_k}} \right)} \right\}}\vspace{0.5ex} \\
  {{v_{{V_m}}}\left( {{t_k}} \right){\rm{ = }}v_{{V_m}}^{real}\left( {{t_k}} \right) + {n_v} \cdot {e^{j{n_\omega }}}}
  \end{array} \label{trans_define}
\end{equation}
where ${n_v}\sim {\rm{N}}\left( {0,\sigma _v^2} \right)$ and ${n_\omega}\sim{\rm{N}}\left( {0,\sigma _\omega^2} \right)$ is the noise of speed and speed orientation that both follows a Gaussian distribution \cite{mendrzik2019enabling}.
The distribution in (\ref{eq_veh_particle}) can now be described as
\begin{equation}
  {{r}_{{V_m}}}\left( {{t_k}} \right) = {{r}_{{V_m}}}\left( {{t_{k - 1}}} \right) + \left( {{{v}_{{V_m}}}\left( {{t_{k - 1}}} \right) + {{v}_{{V_m}}}\left( {{t_k}} \right)} \right) \cdot \frac{T}{2}\label{trans_model}
\end{equation}

\subsection{Weight update}
For CVT particle filter, the observation in (\ref{cvt_weight_upd}) can be calculated by:
\begin{equation}
  \begin{array}{l}
  {\rm{p}}\left( {{Z_{V\left( u \right)}}\left( {{t_k}} \right)\left| {{\rm{CVT}}_u^{\left( a \right)}\left( {{t_k}} \right)} \right.;{\Upsilon _{V\left( u \right)}}} \right)\\
   = \prod\limits_{{m_u} \in V\left( u \right)} {\sum\limits_{j = 1}^{{N_V}} {{\rm{p}}\left( {{Z_{\left( {{m_u},{i_{{m_u}}}} \right)}}\left( {{t_k}} \right)\left| {{\rm{CVT}}_u^{\left( a \right)}\left( {{t_k}} \right),r_{V,{m_u}}^{\left( j \right)}\left( {{t_k}} \right)} \right.} \right)} }
\end{array}\label{cvt_upd_ob}
\end{equation}
where ${Z_{\left( {{m_u},{i_{{m_u}}}} \right)}}\left( {{t_k}} \right){\rm{ = }}\left\{ {{{\widehat \alpha }_{\left( {{m_u},{i_{{m_u}}}} \right)}}\left( {{t_k}} \right),{{\widehat d}_{\left( {{m_u},{i_{{m_u}}}} \right)}}\left( {{t_k}} \right)} \right\}\ \ $ and ${i_{{m_u}}} = {{\mathbbm{1}}_{u,{m_u}}}\left( {{t_k}} \right)$.
Then the estimation for $u$-th CVT based on $j$-th particle in $m_u$-th vehicle particle filter can be denoted as:
\begin{equation}
  \begin{array}{c}
  \widehat r_{{C_u}}^{\left( {{m_u},j,a} \right)}\left( {{t_k}} \right) = r_{{m_u}}^{\left( j \right)}\left( {{t_k}} \right) + \Delta _{{m_u}}^{\left( a \right)} \cdot \overrightarrow R \left( {{{\widehat \alpha }_{\left( {{m_u},{i_{{m_u}}}} \right)}}} \right)\vspace{0.5ex} \\
  \Delta _{{m_u}}^{\left( a \right)} = {\widehat d_{\left( {{m_u},{i_{{m_u}}}} \right)}}\left( {{t_k}} \right) - d_{{C_u}}^{\left( a \right)}\left( {{t_k}} \right)
\end{array}\label{est_mu_j_a4cvt}
\end{equation}
Suppose the error of CVT follows a zero-mean Gaussian distribution, the PDF in (\ref{cvt_upd_ob}) can be denoted as
\begin{equation}
  \begin{array}{*{20}{l}}
  {{\rm{p}}\left( {{Z_{\left( {{m_u},{i_{{m_u}}}} \right)}}\left( {{t_k}} \right)\left| {{\rm{CVT}}_u^{\left( a \right)}\left( {{t_k}} \right);r_{V,{m_u}}^{\left( j \right)}\left( {{t_k}} \right)} \right.} \right)}\\
  { \propto \exp \left\{ { - {{\left( {r_{{C_u}}^{\left( a \right)}\left( {{t_k}} \right) - \hat r_{{C_u}}^{\left( {{m_u},j,a} \right)}\left( {{t_k}} \right)} \right)}^2}} \right\}}
  \end{array}\label{cvt_ob_upd}
\end{equation}

For vehicle particle filter, the observation in (\ref{eq_veh_weight_upd}) is given by:
\begin{equation}
  \begin{array}{l}
{\rm{p}}\left( {{Z_m}\left( {{t_k}} \right)\left| {r_{{V_m}}^{\left( j \right)}\left( {{t_k}} \right);{\Upsilon _{C\left( m \right)}}\left( {{t_k}} \right)} \right.} \right)\\
 = \prod\limits_{{u_m} \in C\left( m \right)} {\sum\limits_{a = 1}^{{N_C}} {p\left( {{Z_{\left( {m,{i_m}} \right)}}\left( {{t_k}} \right)\left| {r_{{V_m}}^{\left( j \right)}\left( {{t_k}} \right);{\rm{CVT}}_{{u_m}}^{\left( a \right)}\left( {{t_k}} \right)} \right.} \right)} }
\end{array} \label{eq_veh_upd}
\end{equation}
where ${Z_{\left( {m,{i_m}} \right)}}\left( {{t_k}} \right){\rm{ = }}\left\{ {{{\widehat \alpha }_{\left( {m,{i_m}} \right)}}\left( {{t_k}} \right),{{\widehat d}_{\left( {m,{i_m}} \right)}}\left( {{t_k}} \right)} \right\}$ and ${i_m} = {{\mathbbm{1}}_{u,m}}\left( {{t_k}} \right)$. Then the estimation for $m$-th vehicle based on $a$-th particle in $u_m$-th CVT particle filter can be denoted as:
\begin{equation}
  \begin{array}{c}
\widehat r_{{V_m}}^{\left( {{u_m},a} \right)}\left( {{t_k}} \right) = r_{{C_{{u_m}}}}^{\left( a \right)}\left( {{t_k}} \right) - {\Delta _{{u_m}}} \cdot \overrightarrow R \left( {{\alpha _{\left( {m,{i_m}} \right)}}\left( {{t_k}} \right)} \right)\vspace{0.5ex} \\
{\Delta _{{u_m}}} = {\widehat d_{\left( {m,{i_m}} \right)}}\left( {{t_k}} \right) - d_{{C_{{u_m}}}}^{\left( a \right)}\left( {{t_k}} \right)
\end{array} \label{veh_ob_upd_particle}
\end{equation}
Then the PDF in (\ref{eq_veh_upd}) can be denoted as:
\begin{equation}
  \begin{array}{l}
{\rm{p}}\left( {{Z_{\left( {m,{i_m}} \right)}}\left( {{t_k}} \right)\left| {r_{{V_m}}^{\left( j \right)}\left( {{t_k}} \right);{\rm{CVT}}_{{u_m}}^{\left( a \right)}\left( {{t_k}} \right)} \right.} \right)\\
 \propto \exp \left\{ { - {{\left( {r_{{V_m}}^{\left( j \right)}\left( {{t_k}} \right) - \widehat r_{{V_m}}^{\left( {{u_m},a} \right)}\left( {{t_k}} \right)} \right)}^2}} \right\}
\end{array} \label{vehicle_ob_upd}
\end{equation}

\subsection{State Estimation}\label{AA}
\noindent The state of CVT cluster $u$ can be estimated as:
\begin{equation}
{\widehat {{\rm{CVT}}}_u}\left( {{t_k}} \right) \approx \sum\limits_{a = 1}^{{N_C}} {w_{{C_u}}^{(a)}\left( {{t_k}} \right) \times {\rm{CVT}}_u^{(a)}\left( {{t_k}} \right)}  \label{eq_cvt_cal}
\end{equation}
The state of vehicle $m$ can be estimated as:
\begin{equation}
{\widehat r_{{V_m}}}\left( {{t_k}} \right) \approx \sum\limits_{j = 1}^{{N_V}} {w_{{V_m}}^{(j)}\left( {{t_k}} \right) \times r_{{V_m}}^{(j)}} \left( {{t_k}} \right)\label{eq_veh_cal}
\end{equation}

\begin{figure*}[hbpt]
  \centering
\includegraphics[width=0.75\textwidth]{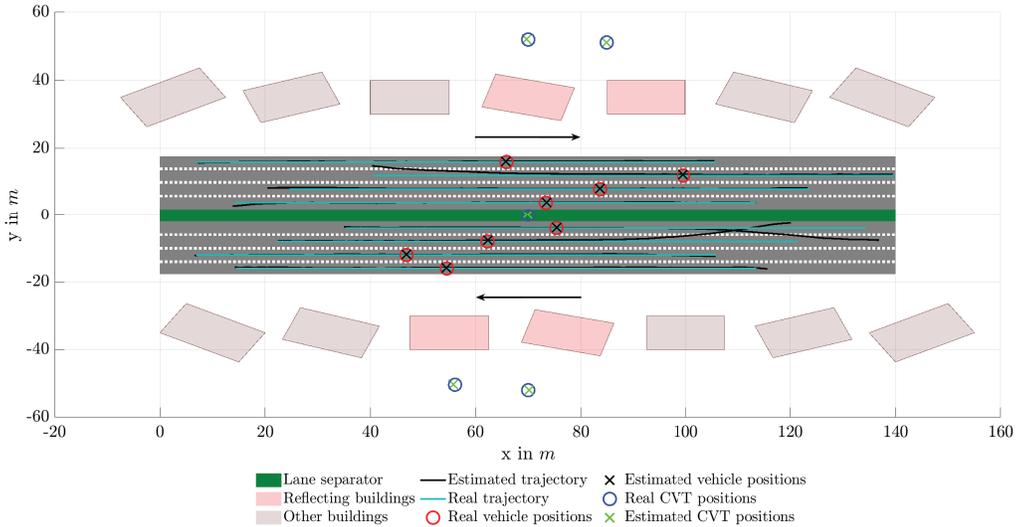}
\caption{Scenario, trajectory and planform. Base station locates at $\left[ {70m,0,8m} \right]$ in the lane separator and the buildings locate beside the road. There are 8 lanes on the road, where vehicles travel along $+x$ direction and $-x$ direction. The real vehicle trajectory and estimated trajectory are shown in blue lines and black lines, respectively.  }
\label{fig_trajectory}
\end{figure*}

\begin{algorithm}[t]
\caption{Team Channel-SLAM}
\LinesNumbered
Initialize the initial particles $r_{{V_m}}^{(j)}\left( {{t_0}} \right)$ from ${r_{{V_m}}}\left( {{t_0}} \right)$
\For {$t = 1:{t_K}$}{
    Detect and track the multi-path \\
    \If{$new\ path\ is\ detected$}{
      Initialize new CVT clusters for the related VTs
    }
    \If {$tracking\ of\ path\ lost$}{
      Delete the corresponding VT from CVT cluster\\
    }
    Update the CVT cluster as (\ref{ap})$\sim$(\ref{CVTI_def})\\
    \For {$u = 1:{N_{C}}\left( {{t_k}} \right)$}{
      \For{$a = 1:{N_u}$}{
        Draw the CVT particles as (\ref{eq_cvt_particle})\\
        Calculate $w_{C_u}^{(a)}\left( {{t_k}} \right)$ as (\ref{cvt_weight_upd}, \ref{cvt_upd_ob}, \ref{est_mu_j_a4cvt}, \ref{cvt_ob_upd})\\
        }
    }
   Resampling the CVT particle weight ${w_{C_u}^{(a)}\left( {{t_k}} \right)}$\\
   Estimate CVT state ${{{\widehat {{\rm{CVT}}}}_u}\left( {{t_k}} \right)}$ as (\ref{eq_cvt_cal})\\
    \For{$m = 1:M$}{
      \For{$j = 1:{N_{{V_m}}}$}{
        Calculate $w_{{V_m}}^{(j)}\left( {{t_k}} \right)$ as (\ref{eq_veh_weight_upd}, \ref{eq_veh_upd}, \ref{veh_ob_upd_particle}, \ref{vehicle_ob_upd})\\
      }
    }
    Resampling the vehicle particle weight $w_{{V_m}}^{(j)}\left( {{t_k}} \right)$\\
    Estimate the vehicle state ${{{\widehat r}_{{V_m}}}\left( {{t_k}} \right)}$ as (\ref{eq_veh_cal})\\
}
\label{code:recentEnd}
\end{algorithm}

\section{Simulation Results}
In this section, a simulation experiment is carried out to test the performance of the proposed algorithm. The simulation is done under a single bounce reflection model and we suppose that there is no scattering and multi-paths are well detected and tracked. Note that the information about position of the base station is not required in this paper.
Distance measurements and angle measurements in (\ref{ob_state}) are corrupted with zero-mean Gaussian noise with standard deviation ${\sigma _d} = 0.2{\rm{m}}$ and ${\sigma _\alpha } = {\rm{1}}\;\deg $ (both for polar angle and azimuth angle) \cite{mendrzik2019enabling}.
The initial position of vehicles are obtained from GPS and its error follows a zero-mean Gaussian distribution with standard deviation ${\sigma _\varepsilon } = 3{\rm{m}}$.
The standard deviation of speed and its orientation noise is ${\sigma _v} = 0.1m/s$ and ${\sigma _\omega } = 0.1\deg /s$. Notice that the Gaussian distribution mentioned above are cut to $2\sigma\  \left( {{\rm{P}}\left( { - 2\sigma  \le X \le 2\sigma } \right) = 0.9544} \right)$ to prevent estremely big errors from Gaussian distribution, which is impossible in practice.
The sampling interval ${t_\delta }$ is $0.1{\rm{s}}$. The number of particles are set as 120 (${N_V} = {N_C} = 120$). We track the movement of vehicles for 100 time slots. The results are based on 200 times simulation run.
\subsection{Trajectory tracking}
The trajectory and simulation scenario are shown in Fig. \ref{fig_trajectory}. The base station locates at $\left[ {70m,0,8m} \right]$ on the lane separator and the buildings are besides the road (the positions of building are unknown). There are totally 8 lanes on the road and the width of each is 4 meters. The initial position uniformly distribute in $\left( {0,40m} \right)$ for $+x$ running vehicles and $\left( {100m,140m} \right)$ for $-x$ running vehicles. The speed of vehicles is 10 $m/s$. The total trajectory is within $\left[ {0,140m} \right]$ range of $x$-axis and $\left[ { - 16m,16m} \right]$ range of $y$-axis on the road. We explore the performance of the algorithm with different number of vehicles. The number of vehicles per unit area is defined as the vehicle density $\rho _V$. So we totally analyze the algorithm when ${\rho _V} = 1,2,4,6,8\ {{{\rm{vehicles}}} \mathord{\left/
 {\vphantom {{{\rm{vehicle}}} {\left( {{\rm{140 \times 32}}{{\rm{m}}^{\rm{2}}}} \right)}}} \right.
 \kern-\nulldelimiterspace} {\left( {{\rm{140 \times 32}}{{\rm{m}}^{\rm{2}}}} \right)}}$.

\begin{figure*}[http]
   \begin{minipage}[t]{0.25\linewidth}
   \captionsetup{justification=centering}
   \centering
   \centerline{\includegraphics[height=4cm,width=4.5cm]{{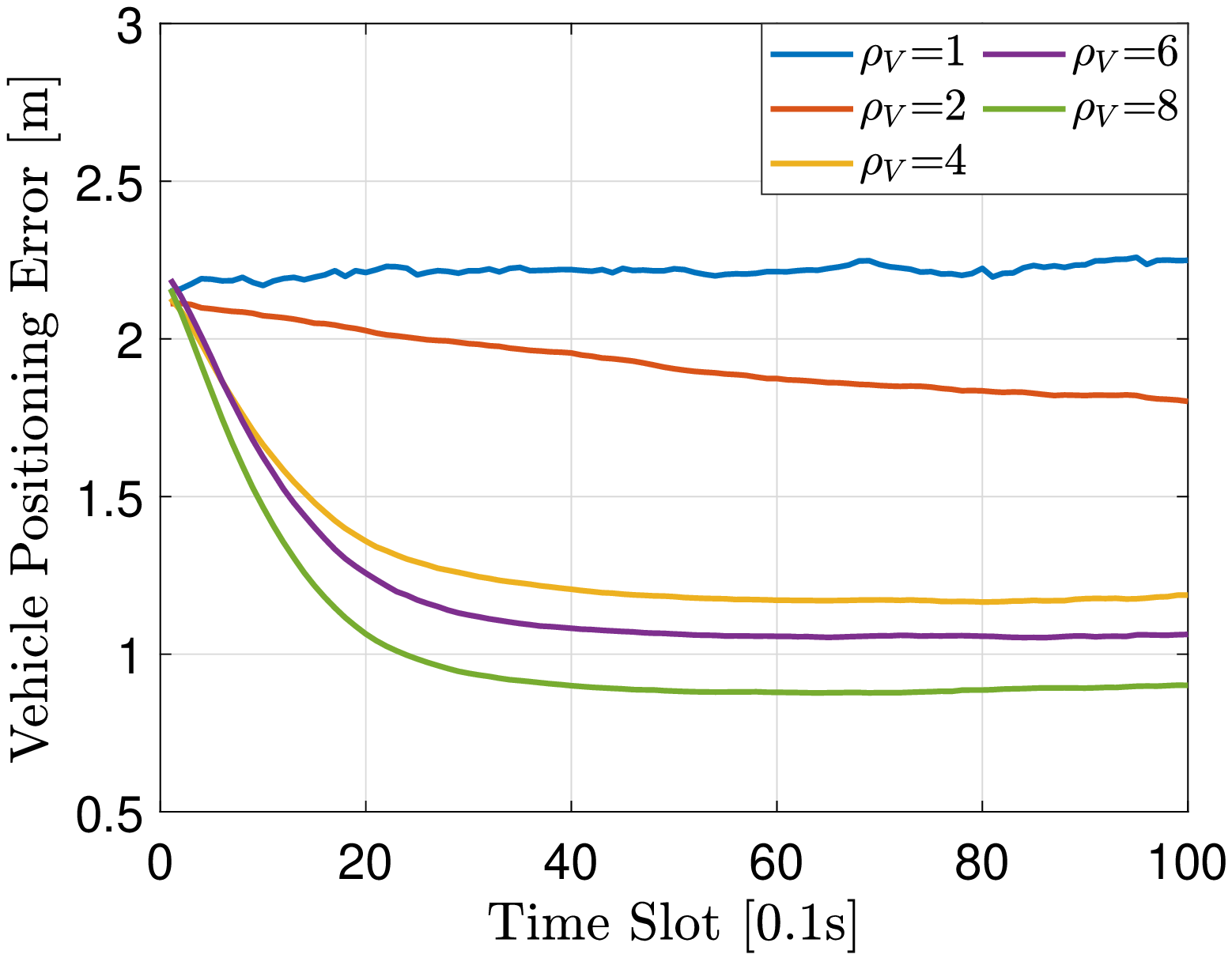}}}
   \caption[justification=centering]{Positioning accuracy expectedly improves over time versus different vehicle densities $\left[ {{{{\rm{vehicle}}} \mathord{\left/
   {\vphantom {{{\rm{vehicle}}} {\left( {{\rm{140 \times 32}}{{\rm{m}}^{\rm{2}}}} \right)}}} \right.
   \kern-\nulldelimiterspace} {\left( {{\rm{140 \times 32}}{{\rm{m}}^{\rm{2}}}} \right)}}} \right]$.}
   \label{fig_error_over_time}
   \end{minipage}%
 \begin{minipage}[t]{0.25\linewidth}
 \captionsetup{justification=centering}
 \centering
 \centerline{\includegraphics[height=4cm,width=4.5cm]{{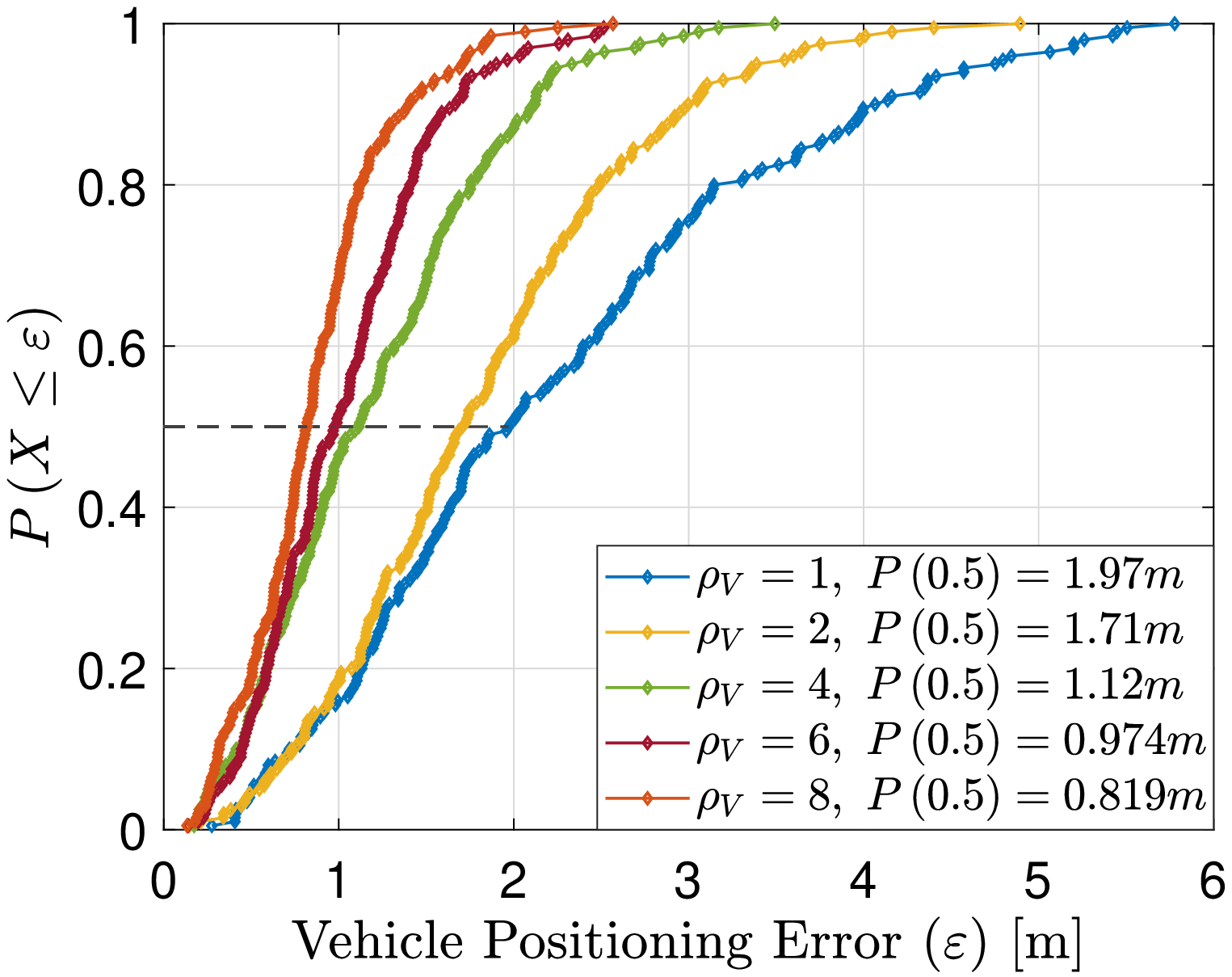}}}
 \caption[justification=centering]{CDF of vehicle positioning accuracy after 100 time slots versus diferent vehicle densities $\left[ {{{{\rm{vehicle}}} \mathord{\left/
 {\vphantom {{{\rm{vehicle}}} {\left( {{\rm{32 \times 140}}{{\rm{m}}^{\rm{2}}}} \right)}}} \right.
 \kern-\nulldelimiterspace} {\left( {{\rm{32 \times 140}}{{\rm{m}}^{\rm{2}}}} \right)}}} \right]$.}
 \label{fig_cdf_error}
 \end{minipage}%
 \begin{minipage}[t]{0.5\linewidth}
   \captionsetup{justification=centering}
 \centering
 \centerline{\includegraphics[height=4cm,width=9cm]{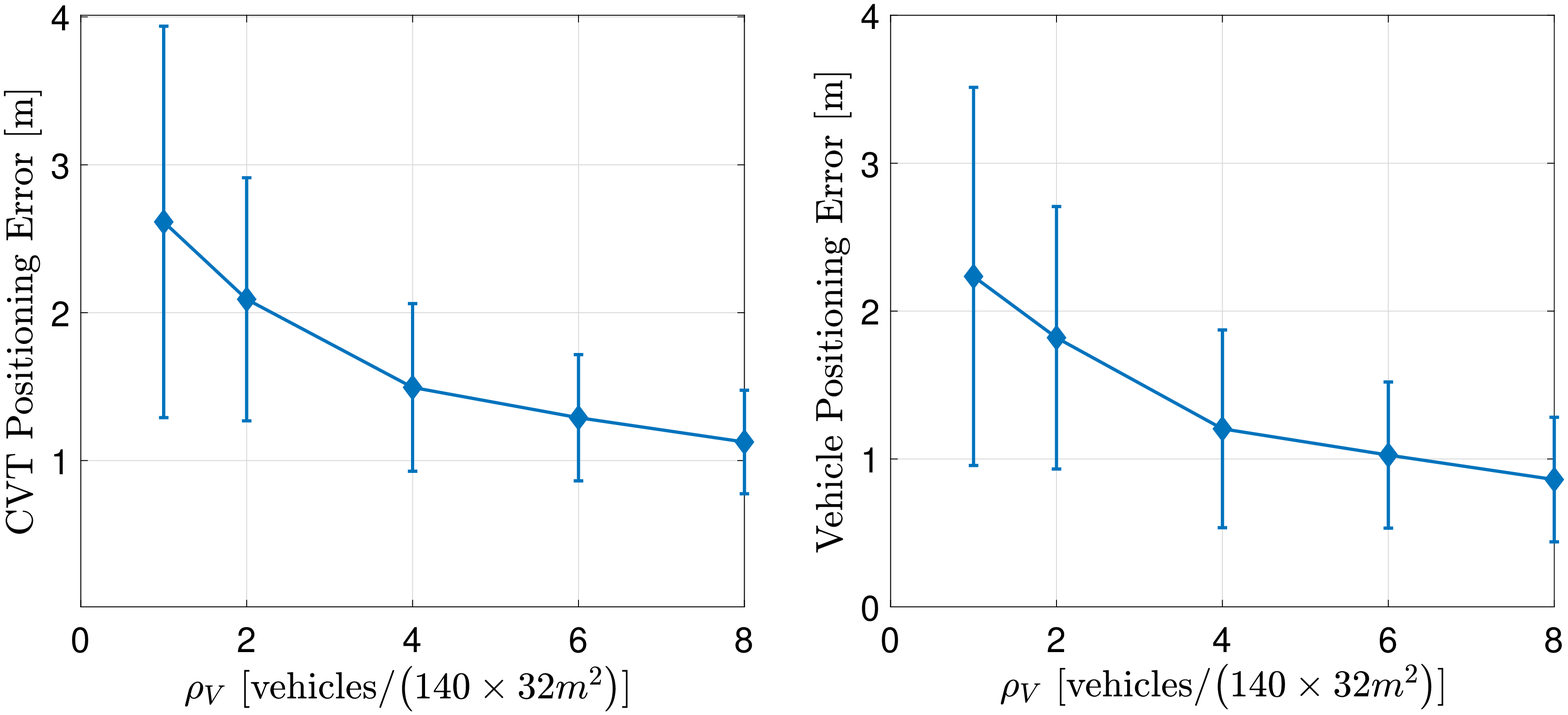}}
  \caption[justification=centering]{Positioning accuracy after 100 time slots expectedly improves with vehicle densities.}
 \label{fig_error_over_density}
\end{minipage}
 \end{figure*}

The figure also shows the trajectory of vehicles when ${\rho _V} = 8\ {{{\rm{vehicle}}} \mathord{\left/
 {\vphantom {{{\rm{vehicles}}} {\left( {{\rm{140 \times 32}}{{\rm{m}}^{\rm{2}}}} \right)}}} \right.
 \kern-\nulldelimiterspace} {\left( {{\rm{140 \times 32}}{{\rm{m}}^{\rm{2}}}} \right)}}$.
 The estimated trajectory and real trajectory are marked by black lines and blue lanes, respectively. Also the real positions and estimated positions for vehicles and CVTs at the 60-th time slot are shown in the figure. We can see that the estimated vehicle trajectory gets gradually close to the real trajectory along with time, where the tendency can also be shown in Fig. \ref{fig_error_over_time}, which means that Team Channel-SLAM can track the vehicle well despite the large initial position errors. A more precise quantitative analysis for errors over time and vehicle density is done in the following subsections.

\subsection{Positioning error over time}
Fig. \ref{fig_error_over_time} shows the positioning error of vehicles over time versus different vehicle densities. The positioning error is high at beginning for the initial positioning error from GPS is large but it converges to a low value over time. This shows that the accumulating observation of CVTs will bring more accurate CVT positioning, which will finally improve the vehicle positioning. But the positioning error doesn't decrease over time in single vehicle situation. This is because in this situation the CVT estimation is based on the relative distance observation (ToA and AoA) and the initial position of just one vehicle.
Thus the error from initial position of that vehicle will completely spread to the CVTs (CVT is demoted to VT in single vehicle situation).
So limited by the given information, the best performance for Team channel-SLAM in single vehicle situation is to keep the initial position error from increasing by exploring the accumulative observations over time. But Team Channel-SLAM is also useful in this situation, because it will eliminate the accumulative error from motion information and also provide a continuous and stable positioning for vehicles in case that the GPS or other positioning sensors get limited by the weather or other hard conditions.
However, when there are multiple vehicles, those CVTs that contain VTs observed by multiple vehicles will get better estimated due to multiple observations, which will improve the positioning of their corresponding vehicles. Also those vehicles will make all of their corresponding CVTs get better estimated in return and then a positive feedback for positioning is spread among all the vehicles and CVTs to converge their particles to real positions, which makes Team Channel-SLAM performs better in multiple vehicle situation.

\subsection{Positioning error over vehicle density}
Fig. \ref{fig_cdf_error} shows the cumulative probability distribution function (CDF) of positioning error after 100 time slots versus different vehicle densities. Fig. \ref{fig_error_over_density} shows the positioning error of CVTs and vehicles after 100 time slots versus different vehicle densities. We can see from the figures, the positioning accuracy expectedly improves over vehicle density for CVT positioning and vehicle positioning. Especially, when ${\rho _V} = 4\ {{{\rm{vehicle}}} \mathord{\left/
 {\vphantom {{{\rm{vehicles}}} {\left( {{\rm{140}} \times {\rm{32}}{{\rm{m}}^{\rm{2}}}} \right)}}} \right.
 \kern-\nulldelimiterspace} {\left( {{\rm{140}} \times {\rm{32}}{{\rm{m}}^{\rm{2}}}} \right)}}$, the algorithm leads to more than ${\rm{40\% }}$ improvement over single vehicle situation. This shows that the increasing number of vehicles will lead to spatial accumulating observation of the CVTs, which will improve the positioning for CVTs and vehicles.

\section{Conclusion}
Team Channel-SLAM models the CVTs based on affinity propagation clustering and then simultaneously estimate the state of CVTs and vehicles through cooperative particle filters. The simulation results show that the algorithm can converge the position of CVTs and vehicles gradually over time and the performance can also be improved when vehicle density increases.
\section*{Acknowledgments}
This work is partially supported by Research on the key technology of seamless handover in cloud-RAN for network assisted automatic driving, NSFC project, under grant 61801047. It is also supported by Beijing Nova Program of Science and Technology under grant Z191100001119028.

\bibliographystyle{IEEEtran}
\bibliographystyle{unsrt}
\bibliography{IEEEabrv,ref}

\end{document}